\documentclass{article}

\usepackage{PRIMEarxiv}

\usepackage[utf8]{inputenc} 
\usepackage[T1]{fontenc}    
\usepackage{hyperref}       
\usepackage{url}            
\usepackage{booktabs}       
\usepackage{amsfonts}       
\usepackage{nicefrac}       
\usepackage{microtype}      
\usepackage{lipsum}         
\usepackage{fancyhdr}       
\usepackage{graphicx}       
\graphicspath{{media/}}     

\usepackage{subcaption}  
\usepackage{multirow}
\usepackage{array}
\usepackage{rotating}
\usepackage{calc}
\usepackage{boldline}
\usepackage{arydshln}
\usepackage{rotating}
\usepackage{caption}
\usepackage{afterpage}
\usepackage{placeins}
\usepackage{floatrow}

\usepackage{amsmath}
\usepackage{algorithm}
\usepackage[noend]{algpseudocode}

\usepackage{xcolor}  

\title{Autonomous Vehicles as a Sensor: Simulating Data Collection Process
}

\author{
  Yunfei Zhang, Mario Ilic, Klaus Bogenberger \\
  Chair of Traffic Engineering and Control \\
  Technical University of Munich (TUM) \\
  Munich, Germany \\
  \texttt{\{yunfei.zhang, mario.ilic, klaus.bogenberger\}@tum.de} \\
}

\begin{document}
\maketitle

\begin{abstract}
Urban traffic state estimation is pivotal in furnishing precise and reliable insights into traffic flow characteristics, thereby enabling efficient traffic management. Traditional traffic estimation methodologies have predominantly hinged on labor-intensive and costly techniques such as loop detectors and floating car data. Nevertheless, the relentless progression in autonomous driving technology has catalyzed an increasing interest in capitalizing on the extensive potential of on-board sensor data, giving rise to a novel concept known as "Autonomous Vehicles as a Sensor" (AVaaS).
This paper innovatively refines the AVaaS concept by simulating the data collection process. We take real-world sensor attributes into account and employ more accurate estimation techniques based on the on-board sensor data. Such data can facilitate the estimation of high-resolution, link-level traffic states and, more extensively, online cluster- and network-level traffic states.
We substantiate the viability of the AVaaS concept through a case study conducted using a real-world traffic simulation in Ingolstadt, Germany. The results attest to the ability of AVaaS in estimating both microscopic (link-level) and macroscopic (cluster- and network-level) traffic states, thereby highlighting the immense potential of the AVaaS concept in effecting precise and reliable traffic state estimation and also further applications.
\end{abstract}


\section{Introduction}

\subsection{Background and Motivation}

Traffic state estimation holds a critical role in the realm of traffic engineering. The advent of a low-latency, precise estimation could prove to be profoundly beneficial for consequent traffic control measures and even long-term planning. In order to enhance the accuracy of traffic state estimation or prediction, the research methodologies have broadened their horizons over recent years to encompass both model-based and data-driven approaches. These methodologies represent a significant shift towards comprehensive and accurate traffic state prediction, underlining the significance of evolving research paradigms in traffic engineering.

The collection of traffic data forms the foundation for effective traffic management, planning, and control. The accuracy and reliability of this data directly impact the effectiveness of traffic state estimation, a crucial component in improving road safety, reducing congestion, and enhancing transportation infrastructure. Traditionally, traffic data collection relied on methods such as manual counting or loop detectors, which are both labor-intensive and often cost-prohibitive. However, with the rise of automated vehicles and smart transportation systems, novel methodologies are emerging, leveraging sensor data from diverse sources such as autonomous vehicles, traffic cameras, and even social media feeds. As we stand on the precipice of a new era in traffic data collection, it is imperative to explore these novel methods and understand their implications for traffic state estimation and management.

Autonomous Vehicles (AVs), furnished with a wide range of sensor technologies such as LiDARs, radars, and cameras, hold the capacity to drastically transform traffic data collection. These AVs serve the dual purpose of being mobile sensors, persistently acquiring real-time traffic data while traversing through complex urban road networks.
 Given that these data are inherently collected for AV functionality, we propose an innovative approach known as Autonomous Vehicles as a Sensor (AVaaS). This model essentially capitalizes on AVs in both motion and parked states to compile valuable data, which can subsequently be applied to traffic state estimation. A distinguishing feature of AVaaS is its expansion beyond the conventional Floating Car Data (FCD) that solely rely on vehicles traversing the street network and only deliver information about the ego-vehicle. AVaaS, on the other hand, benefits from both active and parked vehicles within the network. Additionally, on-board vehicle sensors provide additional information about other road users. Furthermore, since AVs can be centrally regulated by an operator, they can be strategically assigned to specific locations to gather data, particularly when they are not privately owned such as MoD vehicles.

 By harnessing data from AVs, we can broaden the range of traffic state estimation to encompass both microscopic and macroscopic scales. On the microscopic scale, AVs can estimate traffic states such as relative flow and link travel time for individual road segments. Concurrently, AVs furnish invaluable data that aids in calculating traffic parameters like average speed, traffic volume, speed, and traffic density at the macroscopic network level.

In this study, we aim to unravel the potential of leveraging AVs as sensors for traffic state estimation. By simulating the data collection process using real sensor attributes and applying the concept of AVaaS to a real-world traffic network of the city of Ingolstadt, Germany, we seek to find its potentials in networl-level traffic state estimation.

\subsection{Related Work}

The concept of using traffic data from moving vehicles for traffic engineering and control dates back to 1954 when ~\cite{wardrop1954method} proposed the moving observer (MO) method. This method estimated speed and traffic flow based on manual observations of surrounding traffic in both driving directions, including the number of vehicles overtaking or being overtaken, and the count of oncoming vehicles~\cite{wardrop1954method}. 
Some follow-up methods demonstrated its feasibility in estimating traffic states \cite{Mulligan&Nicholson2002} or practical implementation for traffic management \cite{Czogalla&Naumann2007, wolf2008floating}, which are summarized in the following Table \ref{tab:Literature}.

\begin{table}
    \caption{Literature overview of simulation studies on the moving observer method}\label{tab:Literature}
    \centering
    \rotatebox{90}{
    \begin{tabular}{l|c!{\vrule width 1.5pt}c|c!{\vrule width 1.5pt}c|c!{\vrule width 1.5pt}c|c|c|c!{\vrule width 1.5pt}c|c|c|c|c}
        \multirow{14}{*}{Source} & \multirow{14}{*}{\centering Year} & \multicolumn{2}{c!{\vrule width 1.5pt}}{Observer Vehicle} & \multicolumn{2}{c!{\vrule width 1.5pt}}{Direction} & \multicolumn{4}{c!{\vrule width 1.5pt}}{Recorded Traffic Parameters} & \multicolumn{5}{c}{Calculated Traffic Parameters} \\
        \cline{3-15} 
                   &  & \multirow{13}{*}{\centering \rotatebox{270}{Passenger Car}} & \multirow{13}{*}{\centering \rotatebox{270}{Public Transportation}} & \multirow{13}{*}{\centering \rotatebox{270}{Same-Directional}} & \multirow{13}{*}{\centering \rotatebox{270}{Oncoming Traffic}} & \multirow{13}{*}{\centering \rotatebox{270}{Relative Flow (same-directional)}} & \multirow{13}{*}{\centering \rotatebox{270}{Relative Flow (oncoming traffic)}} & \multirow{13}{*}{\centering \rotatebox{270}{Link Travel Time}} & \multirow{13}{*}{\centering \rotatebox{270}{Avg. Speed (of detected vehicles)}} & \multirow{13}{*}{\centering \rotatebox{270}{Traffic Volume}} & \multirow{13}{*}{\centering \rotatebox{270}{Speed}} & \multirow{13}{1.0cm}{\centering \rotatebox{270}{Traffic Density}} & \multirow{13}{1.0cm}{\centering \rotatebox{270}{Travel Time}} & \multirow{13}{*}{\centering \rotatebox{270}{Queue Lengths at Intersections}} \\
         &  &  &  &  &  &  &  &  &  &  &  &  &  & \\
         &  &  &  &  &  &  &  &  &  &  &  &  &  & \\
         &  &  &  &  &  &  &  &  &  &  &  &  &  & \\
         &  &  &  &  &  &  &  &  &  &  &  &  &  & \\
         &  &  &  &  &  &  &  &  &  &  &  &  &  & \\
         &  &  &  &  &  &  &  &  &  &  &  &  &  & \\
         &  &  &  &  &  &  &  &  &  &  &  &  &  & \\
         &  &  &  &  &  &  &  &  &  &  &  &  &  & \\
         &  &  &  &  &  &  &  &  &  &  &  &  &  & \\
         &  &  &  &  &  &  &  &  &  &  &  &  &  & \\
         &  &  &  &  &  &  &  &  &  &  &  &  &  & \\
         &  &  &  &  &  &  &  &  &  &  &  &  &  & \\ \hlineB{4}
        \text{\cite{Mulligan&Nicholson2002}} & 2002 & x &  &  & x &  & x & x & x & x &  &  &  &  \\ \hline
        \text{\cite{Czogalla&Naumann2007}} & 2007 &  & x & x & x & x & x &  & x &  &  &  & x &  \\ \hline
        \text{\cite{wolf2008floating}} & 2008 &  & x &  & x &  & x &  &  & x &  &  &  &  \\ \hline
        \text{\cite{kuhnel2009evaluation}} & 2009 & x & x &  & x &  & x & x & x & x &  &  &  &  \\ \hline
        \text{\cite{Florin&Olariu2016}} & 2016 & x &  & x &  & x &  & x &  & x & x & x &  &  \\ \hline
        \text{\cite{schafer2017bewegte}} & 2017 & x &  & x & x & x & x &  &  & x & x &  & x &  \\ \hline
        \text{\cite{vanErpetal2018}} & 2018 & x &  & x &  & x &  &  &  & x &  & x &  &  \\ \hline
        \text{\cite{Guerrierietal2019}} & 2019 & x &  & x & x & x & x & x &  & x & x & x &  &  \\ \hline
        \text{\cite{vanErpetal2019}} & 2019 & x &  & x &  & x &  &  &  & x &  & x &  &  \\ \hline
        \text{\cite{Langeretal2020}} & 2020 & x &  &  & x &  & x &  &  &  &  &  &  & x \\ \hline
        \text{\cite{vanErp2020}} & 2020 & x &  & x & x & x &  &  &  & x &  & x &  &  \\ \hline
        \text{\cite{Ma&Qian2021}} & 2021 & x &  & x & x & x & x & x &  & x & x & x &  &  \\ \hline
        \text{\cite{Florin&Olariu2023}} & 2023 & x &  & x & x & x & x &  &  &  &  & x &  &  \\

    \end{tabular}
    }
\end{table}

When it comes to the calculated traffic parameters, the calculation of instantaneous link-based fundamental traffic flow variables, such as the traffic volume, the speed or the traffic density is the primary goal of most of the investigated literature. \cite{Czogalla&Naumann2007} as well as ~\cite{schafer2017bewegte, schaferverkehrszustandsschatzung} additionally investigated the calculation of link-based travel times from the recorded traffic parameters, while ~\cite{Langeretal2020} obtained queue lengths at intersections from relative flow data collected from oncoming traffic flows.

Based on these fundamental studies, the following studies extend the MO method into different aspects. From the aspect of transportation modes, \cite{Czogalla&Naumann2007, kuhnel2009evaluation} investigated the impact of different types of moving observers on traffic flow estimation, considering both passenger cars and public transportation.
From the aspect of estimation methods, \cite{Florin&Olariu2016} introduced a data-driven approach using machine learning to predict traffic flow characteristics based on continuous data streams from mobile sensors. \cite{Guerrierietal2019} focused on data fusion techniques to enhance traffic data accuracy and completeness through multiple moving observers. \cite{Ma&Qian2021} integrated machine learning techniques with moving observers to improve prediction accuracy. 

With the advancement in vehicle sensor technology of the last decades, the data acquisition of surrounding traffic conditions can now be performed by on-board sensors of connected vehicles, providing extended floating car data (xFCD) for the use in Intelligent Transportation System applications.
\cite{schafer2017bewegte, schaferverkehrszustandsschatzung} emphasized the integration of moving observers in urban traffic management systems, particularly in the context of connected and autonomous vehicles. \cite{vanErpetal2018} compared traffic state estimation methods based on floating cars and moving observers, providing insights into their performance.

Overall, the simulation studies have primarily focused on the investigation of the feasibility of using the Moving Observer (MO) method to collect traffic data. These studies have provided valuable insights into potential applications and limitations of MO data but have also exposed gaps: 1) The current literature emphasizes data collection during motion, with little exploration of stationary scenarios, and 2) The studies have been more concerned with collecting traffic data rather than its application in traffic engineering and management.

To address these research gaps, we introduced our original concept, Autonomous Vehicles as a Sensor (AVaaS)~\cite{zhang2023novel}. AVaaS expands on existing research in two main ways: Firstly, it proposes the use of \textit{parked observers (PO)} as stationary detectors to gather traffic data at fixed locations. Secondly, it leverages this collected data for link-based and network-based traffic state estimation, represented by the (macroscopic) fundamental diagram.
Current research on network-level traffic state estimation primarily relies on loop detector data or floating car data (FCD). AVaaS, however, opens new avenues by utilizing on-board sensors like cameras, LiDAR, and radar in autonomous vehicles (AVs) to provide extended FCD (xFCD), offering additional insights into the vehicle's surroundings. This enhancement over conventional methods enables a diverse range of applications, including the monitoring of intersections for queue lengths and pedestrian counts, the estimation of traffic flows, speeds, and densities on road sections, and the detection and assessment of incidents in opposite driving directions~\cite{banerjee2018online}.

Considering Autonomous Mobility-on-Demand (AMoD) services, AVaaS can be extended to a more general vehicle assignment strategy. Specifically, idle AVs can be assigned to certain places of the network for traffic data collection, either as MO or PO. This approach has been shown to effectively capture traffic conditions, as demonstrated by~\cite{zhang2022temporal} by using GPS trajectory data from ride-hailing vehicles to model traffic flow using a two-fluid model.
Figure~\ref{fig:moving_observer_mod} illustrates how the AVaaS idea is applied to AMoD services in a generalized way. Each AMoD vehicle collects local information from various modes for both road sections and intersections. The collected data is then sent to the AMoD operator to estimate network-level traffic states. Finally, the operator can decide to dispatch (indicated by the blue arrows in the figure) an idle vehicle to a specific location (indicated by the grey box with blue outline) to collect local traffic data. By utilizing AVaaS to collect traffic data from areas where no traffic state information is available, the AMoD operator can fill the gap of information and improve the network-wide traffic state estimation.

\textbf{\begin{figure*}[!h]
    \centering
    \includegraphics[width=1\linewidth]{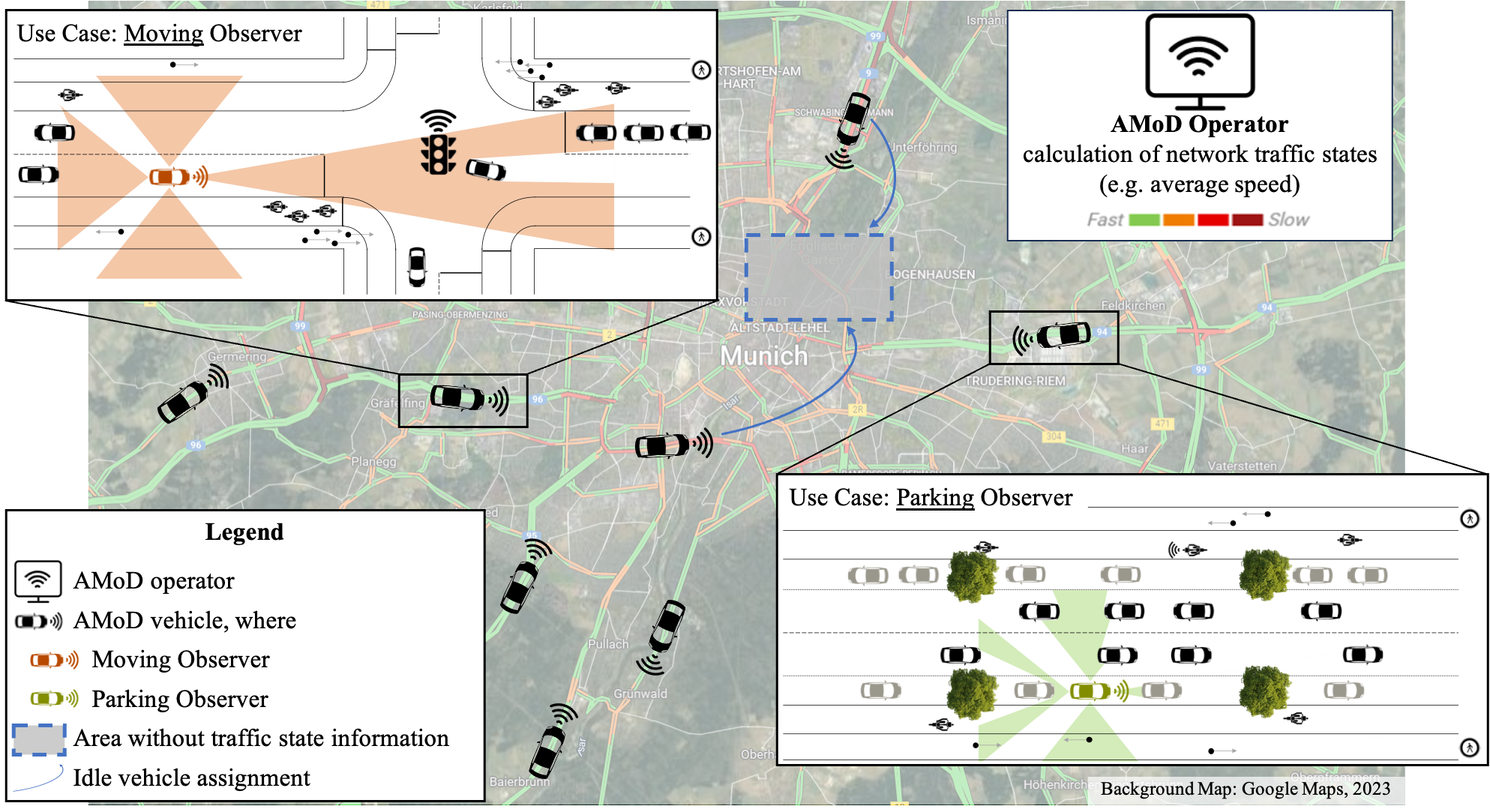}
    \caption{The application of AVaaS on AMoD services}
    \label{fig:moving_observer_mod}
\end{figure*}}

However, since fully autonomous vehicles have not yet been deployed on the large-scale, we need to simulate the data collection process for AVs to study its effects. There are several ways to achieve this: in the original AVaaS paper, we proposed a simple lane- and distance-based method \cite{zhang2023novel}; a more complicated 3D vision-based method was studied by \cite{gerner2023enhancing}.
To evaluate the simulation process, there also exists several indicators \cite{huber2014new, breitenberger2004extended}.

\subsection{Research Gap and Research Questions}\label{subsec:rq}
The major limitation of the initial paper \cite{zhang2023novel} is that only a grid network and synthesized demand have been investigated in the case study and only macroscopic fundamental diagrams (MFDs) \cite{daganzo2007urban} have been compared with ground truth. Besides, the detection and estimation methods are simplified using simple lane- and distance-based methods. To tackle these limitations, we derive the following research questions for this paper:

\begin{enumerate}
    \item Is the AVaaS concept able to estimate the traffic states in a real road network?
    \item How does the AVaaS concept perform in different levels of traffic state estimation?
    \item Can detection and estimation methods be further improved?
\end{enumerate}

\subsection{Contribution}

The key contributions of this research are as follows:

Simulating data collection process using real-world detection factors: In the pursuit of a more realistic and understanding of traffic flow, we simulates the data collection process using real-world detection factors. 
Unlike the preceding conceptual paper~\cite{zhang2023novel} that have relied on simplified assumptions, we incorporates actual range limitations and other intricate details such as detection angles from real-world sensors. 
By meticulously replicating these factors, we pursue to mimic the data collection process for fully autonomous vehicles in the future.

Real network as a case study:
Another limitation in the previous phase~\cite{zhang2023novel} is that we used a grid network and synthesized demand profile.
In this paper, we validate our idea in a real network used as a case study. Different from homogeneous grid network, we encounter the complexities and heterogeneity of real-world traffic dynamics. 
The study brings insights of applying this method in a real application and thus, encouraging the further study of methodology to tackle the cluster- or even network-level challenges.

Different levels of traffic state estimation:
Finally, our research breaks new ground by introducing different levels of traffic state estimation: starting from link-level traffic state estimation, 
we aggregate the traffic states to the cluster and then network level. Recognizing that, further applications such as routing~\cite{zhang2023network} can be achieved for various granularities, we have devised methods to analyze it at both the microscopic and macroscopic levels. The microscopic analysis inspects individual lanes including density, speed, and vehicles entered or left, while the macroscopic analysis firstly characterizes the road geography and then overviews the network performances.
This multi-granularity analysis provides further research opportunities or possibilities for further clustering or network-level traffic state estimation.

Through these key contributions, our research aims to address the challenges of traffic state estimation using AVaaS and shed lights on further data-driven traffic management strategies. By simulating real-world data collection, utilizing real network scenarios, and extending traffic state estimation to different levels, this paper seeks to bring AVaaS from a concept to applications in the future.

\section{Methodology}

Originated from the methodology we introduced in \cite{zhang2023novel}, we modify it based on the new detection and estimation range.

\subsection{Surrounding Vehicle Detection}
Both the Moving Observer (MO) concept and the Autonomous Vehicle as a Service (AVaaS) idea are rooted in the detection of surrounding vehicles, a fundamental step in estimating traffic states. In real-world scenarios, an ego vehicle employs a range of sensors to detect its surroundings, utilizing technologies such as Vehicle-to-Vehicle (V2V) and Vehicle-to-Infrastructure (V2I) communication. The proposed methodology hinges on simulation-based detection methods, developed from simulation results where all vehicle movements are fully observable. Figure~\ref{fig:moving_observer_mod} illustrates the two considered use cases for the detection process: the \textit{Moving Observer} and the \textit{Parking Observer}.

In our conceptual paper \cite{zhang2023novel}, we explore two interrelated concepts: lane-based detection and distance-based detection. Lane-based detection pinpoints vehicles within the same, close, or opposite lanes as the ego vehicle, including only these in subsequent calculations. Following this, distance-based detection computes the Euclidean distance between the ego vehicle and the previously identified nearby vehicles. A vehicle's detection is confirmed only if the distance lies within specific thresholds (detection ranges) $d_l^{f/b}$ for lane $l$ in both forward $f$ and backward $b$ directions. Importantly, these thresholds vary based on the vehicle's position relative to the ego vehicle, as visibility may be obstructed by other vehicles, median strips, or additional obstacles.

It should be noted, however, that even within the detection range, objects might be sensed with noise and their detection could be easily disrupted by closer obstructions. Therefore, we define a more refined detection area in Figure~\ref{fig:detection_zone}, where $w$ represents the horizontal detection width, and $l$ denotes the vertical direction. According to current sensor specifications and real driving conditions, the detection ranges are doubled for the front and left directions compared to the back and right.

For simplification, we previously used a uniform distance of $50$ meters for detection in both forward and backward directions \cite{zhang2023novel}. In this paper, we apply more realistic sensor parameters summarized in Table ~\ref{tab:detector_parameters}, focusing on long-range sensors and aligned with the latest technologies \cite{continental_sensor2023}. We summarize the fundamental parameters of environment sensors as provided by Continental \cite{Continental}, reflecting the consideration of both accuracy and reliability in vehicle detection.

\textbf{\begin{figure*}[!h]
    \centering
    \includegraphics[width=0.3\linewidth]{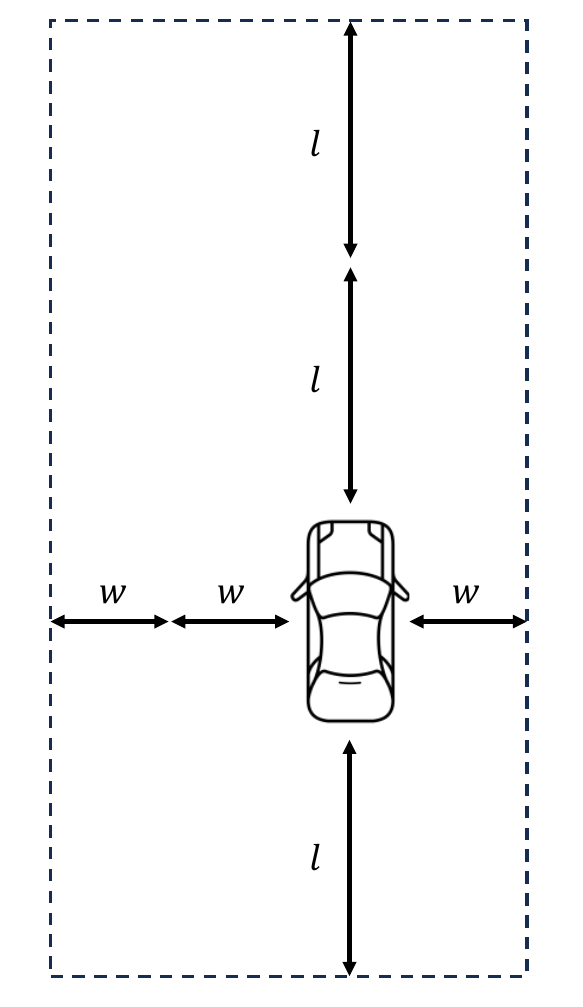}
    \caption{Detection zone of an ego AV}
    \label{fig:detection_zone}
\end{figure*}}

\begin{table}[]
\centering
\caption{Sensor specifications}\label{tab:detector_parameters}
\begin{tabular}{clll}
\hline
Sensor & Type & Range & Field of view \\ \hline
Camera & Mono & & horizontal up to 125°; vertical up to 60° \\
Camera & Rear view & meters & \\
Camera & Surround View & & 360° view \\
Radar & Surround & >180 m & $\pm$ 90° detection / $\pm$ 75° measurement \\
Radar & Long range & 210 - 300 m & $\pm$ 50 - 60° \\
Radar & Short range & 100 m & $\pm$ 90° detection / $\pm$ 75° measurement \\
LiDAR & High resoluton 3D &  & 120° $\times$ 30° \\
LiDAR & Long range & 1000 m & 128° $\times$ 28° \\
LiDAR & Short range & 30 m & \\
\hline
\end{tabular}
\end{table}

After a vehicle is detected, the ego vehicle will compute the distance between the vehicles, determine their current lanes, and measure their respective speeds. These data are then utilized for traffic state estimation purposes. 

\subsection{Traffic State Estimation}

We separate our traffic state estimation methods into two types: microscopic and macroscopic. We first start with the microscopic level estimation and calculate the link-level density and speed. Then, we aggregate these values for cluster- or network-level estimation.

Our approach for network traffic state estimation leverages link-level traffic estimations obtained from AVaaS vehicles. Specifically, we use the macroscopic fundamental diagram (MFD), a widely-used model in traffic engineering introduced by Daganzo~\cite{daganzo2007urban}, to capture the cluster- and network-level traffic state. To estimate the MFD, we need to obtain average densities and flows over a certain period of time. However, these two variables are estimated differently for the simulation-based ground truth and the AVaaS-based approach.

\subsubsection{Ground Truth}

For the simulation-based ground truth, all traffic states are readily available. To aggregate the data, an aggregation level of $t_{agg}$ [s] is defined. The average density $k_{i,ts}$ [veh/km] and the number of entering vehicles $n_{veh}$ [veh] for each lane $i$ during each interval with start time $ts$ are output from the simulation. Therefore, the flow $q_{i,ts}$ [veh/h] can be estimated by:

\begin{equation}
    q_{i,ts} = \frac{n_{veh}*3600}{t_{agg}}
\end{equation}

Subsequently, the mean density $k_{avg,ts}$ and mean flow $q_{avg,ts}$ for a certain aggregation period starting at $ts$ can be averaged among all $n$ lanes:

\begin{equation}
    k_{avg,ts} = \frac{\sum_{i=1}^{n} k_{i,ts}}{n}, \quad q_{avg,ts} = \frac{\sum_{i=1}^{n} q_{i,ts}}{n}
\end{equation}  \newline

\subsubsection{AVaaS Approach}

In the context of the proposed AVaaS approach, the traffic state estimation poses a unique challenge due to the lack of link-level traffic flow data. Consequently, we estimate the traffic flow by the fundamental relationship $q=k*v$. As previously described, we consider two distinct estimation techniques, \textit{Moving Observers (MO)} and \textit{Parking Observers (PO)}, which allows us to vary the estimation based on the specific observer type.

\underline{Moving Observers:} Same as for the calculation of the ground truth, an aggregation level $t_{agg}$ [s] is defined with a start time $ts$. Considering the number of total vehicles at this period $n_{total, ts}$ and a certain penetration rate $p_r (\%)$, we randomly choose $n_{mo} = \text{Round}(n_{total, ts}*p_r/100)$ vehicles as MO, which is at least 1 vehicle per period. 

Starting from link leve, for each observer $j$ and each time step $t \in \{ts, ts+1, \ldots, ts+t_{agg}-1\}$, the number of detected vehicles $n_{det, j, t}$ in a total of $m$ lanes, the detected speed $v_{k, j, t}$ of a surrounding vehicle $k \in \{1, 2, \ldots, n_{det, j, t}\}$ as well as the ego-speed $v_{j, t}$ of the MO $j$ is returned. Subsequently, the traffic density $k_{j,t}^*$, estimated by the MO $j$ at time step $t$ is calculated by:

\begin{equation}\label{eq:est_k}
    k_{j,t}^* = \frac{(1+n_{det, j, t})*1000}{\sum_{l=1}^{m} (d_l^{f}+d_l^{b})}
\end{equation}

where $d_l^{f/b}$ stands for the detection range for a lane $l$ for both forward $f$ and backward $b$ directions. In our current settings of detection zone, $d_l^{f} = 2*d_l^{b}$.

The speed $v_{j,t}^*$ estimated by the MO $j$ at time step $t$ is then calculated by: 

\begin{equation}\label{eq:est_v}
    v_{j,t}^* = \frac{\sum_{k=1}^{n_{det, j, t}} v_{k, j, t} + v_{j, t}}{n_{det, j, t}+1}
\end{equation}

Subsequently, the traffic flow estimated by the MO $j$ at time step $t$ is calculated by:

\begin{equation}\label{eq:est_q}
    q_{j,t}^* = k_{j,t}^* * v_{j,t}^*
\end{equation}

For the cluster- and network-level estimation, we average the estimated densities, speeds, and flows among all MOs $j$ and time steps $t$ in the respective aggregation level $t_{agg}$.
Equation (\ref{eq:est_avg_k}) shows the calculation of the mean density for the aggregation period starting at $t_{s}$, while the aggregation of speeds and flows follows the same procedure.

\begin{equation}\label{eq:est_avg_k}
    k_{avg,ts}^* = \frac{\sum_{j=1}^{n_{mo}} ((\sum_{t=t_{s}}^{ts+t_{agg}-1} k_{j,t}^*)/t_{agg})}{n_{mo}}
\end{equation}

\underline{Parking Observers:} Again, considering the number of total vehicles at this period $n_{total, ts}$ and a certain penetration rate $p_r (\%)$, we randomly choose $n_{po} = \text{Round}(n_{total, ts}*p_r/100)$ vehicles as PO, which is at least 1 vehicle per period. For the estimation of traffic flow variables, the only difference is that POs themselves are not counted in the density and speed calculations, as they are not part of the traffic flow. Therefore, originated from (\ref{eq:est_k}) and (\ref{eq:est_v}), we conclude for the calculation of densities and speeds estimated by the MO $j$ at time step $t$:

\begin{equation}\label{eq:est_k_po}
    k_{j,t}^{**} = \frac{n_{det, j, t}*1000}{\sum_{l=1}^{m} (d_l^{f}+d_l^{b})}
\end{equation}

\begin{equation}\label{eq:est_v_po}
    v_{j,t}^{**} = \frac{\sum_{k=1}^{n_{det, j, t}} v_{k, j, t}}{n_{det, j, t}}
\end{equation}

The remaining averaging process to obtain mean densities, speeds and flows follows the same procedure as described with (\ref{eq:est_avg_k}).
Note that in the case study we will always use half vehicles as MOs and the other half as POs so that the proposed AVaaS method is different from XFCD methods.

\section{Case study}
In this section, we present a case study in a city-size network with realistic traffic demands. Firstly, we briefly describe the setup of SUMO simulation; then, the results of SUMO simulation are presented with the analysis of network characteristics; finally, we present the results from traffic state estimation using AVaaS, from both microscopic and macroscopic levels.

\subsection{Description}

\begin{figure}[!h]
  \centering
  \begin{subfigure}{0.39\textwidth}
    \includegraphics[width=\linewidth]{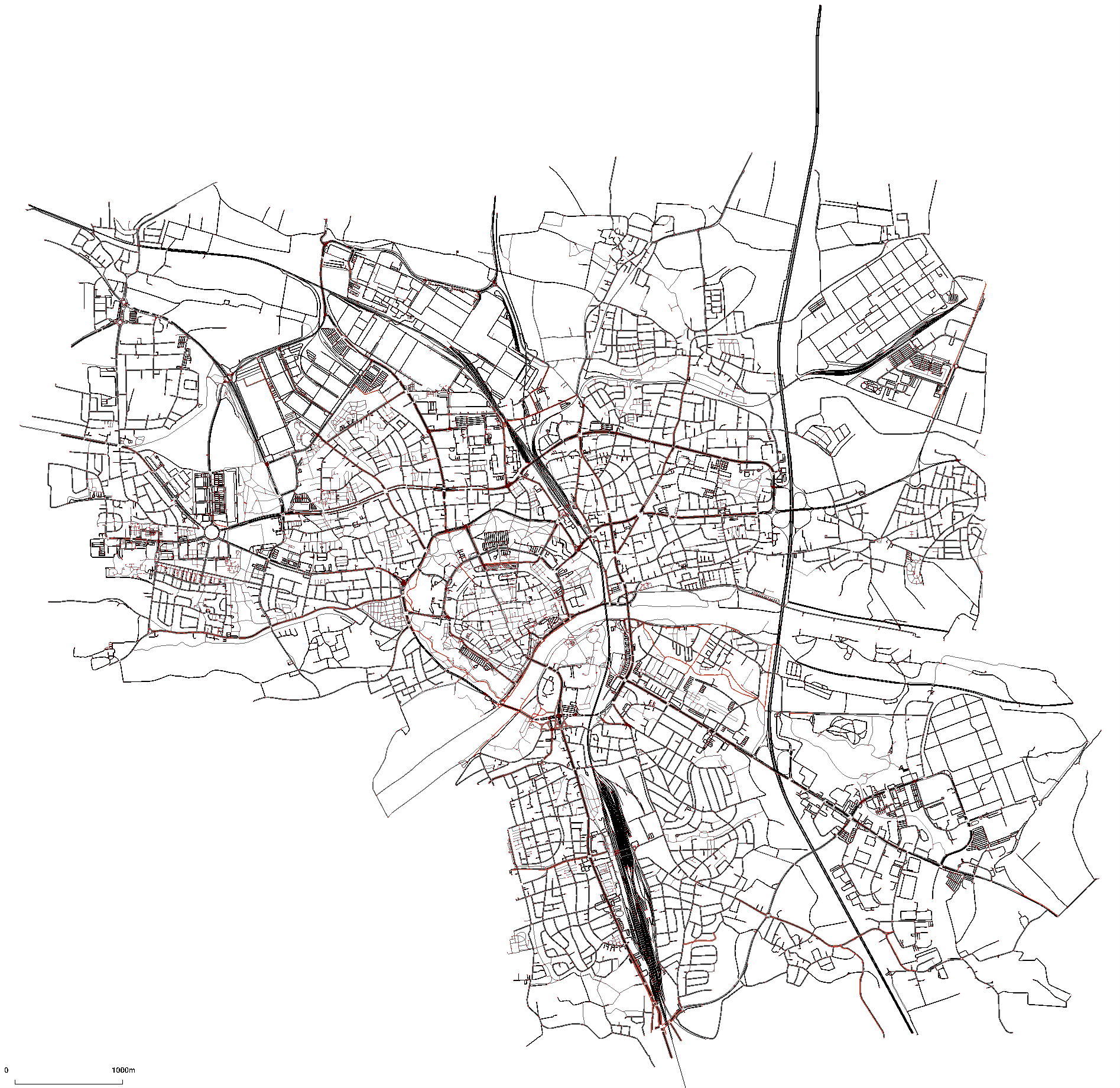}
    \caption{The SUMO road network used in the case study (Ingolstadt, Germany)}
    \label{fig:sumo_ingolstadt_network}
  \end{subfigure}
  \hfill
  \begin{subfigure}{0.59\textwidth}
    \includegraphics[width=\linewidth]{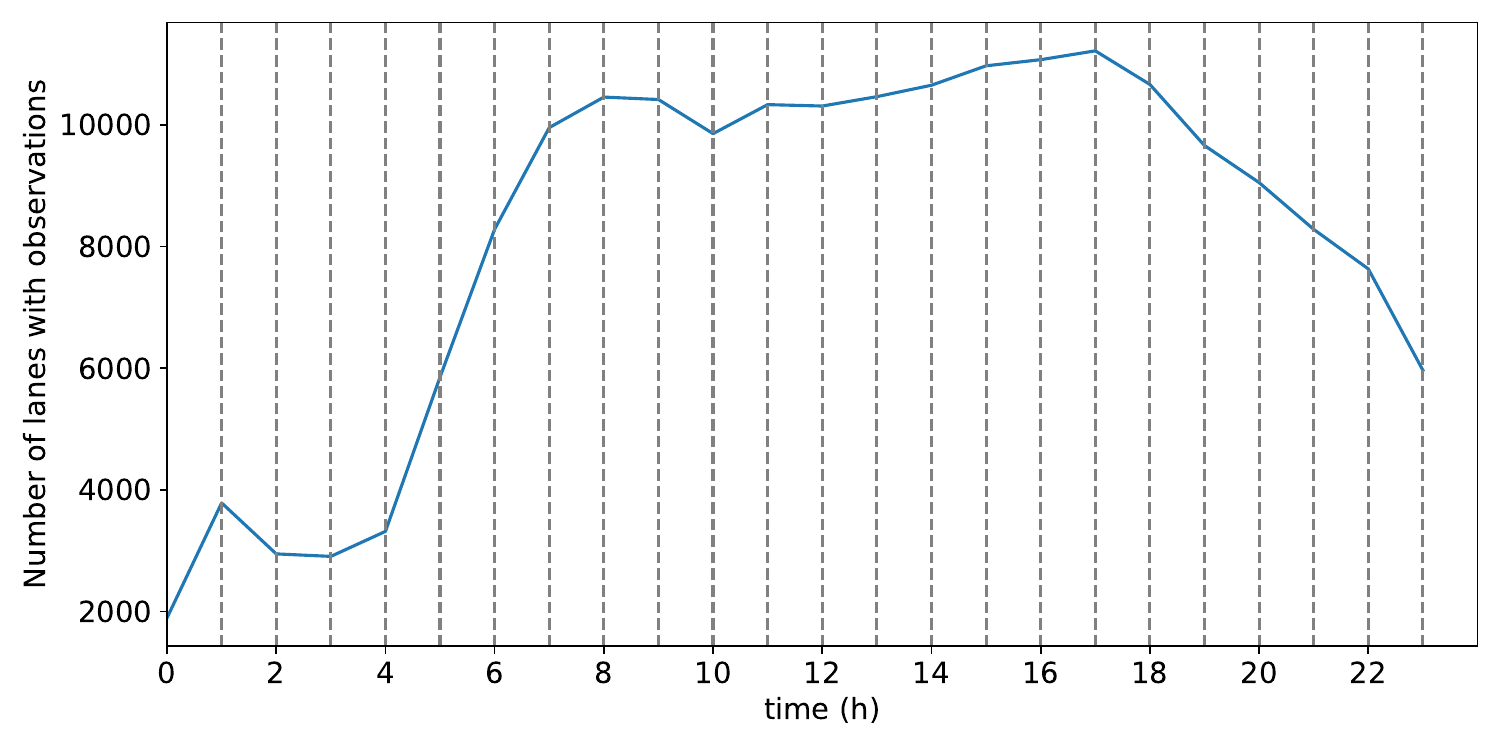}
    \caption{Number of lanes with observations for 24-hour SUMO simulation}
    \label{fig:sumo_lane_distribution_24h}
  \end{subfigure}

  \caption{SUMO simulation in Ingolstadt, Germany for 24 hours}
  \label{fig:simulation_description}
\end{figure}

The case study under consideration utilizes the network from Ingolstadt, Germany, a Bavarian city with an approximate population of 143,000 as of June 2023~\cite{ingolstadt-website2023}. The infrastructure data, including features such as traffic signals, was extracted from an open-source GitHub repository~\cite{sumo-ingolstadt2023}. The Ingolstadt network encompasses approximately 12,000 nodes and 28,000 edges, comprising an assortment of urban, rural, and highway roads as delineated in Figure~\ref{fig:sumo_ingolstadt_network}. Notably, prior calibrations have already been performed on this network, including map association, traffic light program emulation, and traffic flow calibration, thereby increasing its validity for our study~\cite{harth2021automated}.

The demand is also calibrated based on a two-step procedure. Firstly, an initial OD matrix was generated based on socio-demographic statistics. Secondly, these data are calibrated given the traffic counts. Drivers' behaviours are also calibrated based on vehicles trajectory data~\cite{langer2021calibration}.
The initial simulation lasts for a whole-day (24-hour), which we also run and aggregate the results to draw the MFD. However, to evaluate the idea of AVaaS, we only use the morning peak from 6 am to 11 am because the whole simulation will be too huge at the city level.

\subsection{SUMO Simulation and Network Characteristics}

In the simulation, we use $5$ minutes as the aggregation level ($t_{agg}=300 [s]$) to keep the balance between efficient results and fine temporal resolution. Note that we also evaluate the traffic states for lanes instead of edges because edges have different numbers of lanes, which is normally dependent on their road class.

A common fact for a city-level simulation is that the demand is imbalanced both temporarily and spatially: most of vehicles concentrate in the city center during the peak hours. Figure~\ref{fig:sumo_lane_distribution_24h} visualizes the number of lanes with observations over the whole simulation. Considering in total 28,000 edges, which contains at least one lane, there are just over 10,000 lanes with observations with observations during the 5-minute interval. This brings some challenges to our traffic state estimation as we randomly select vehicles as moving observers and parking observers for the current stage.

The MFD for 24-hour simulation is shown in Figure~\ref{fig:sumo_mfd_24h}. We conclude that it is far away from the capacity because Ingolstadt is not a city with heavy traffic. Therefore, it will be interesting to study the case with congestion situations in the future.

\textbf{\begin{figure*}[!h]
    \centering
    \includegraphics[width=0.5\linewidth]{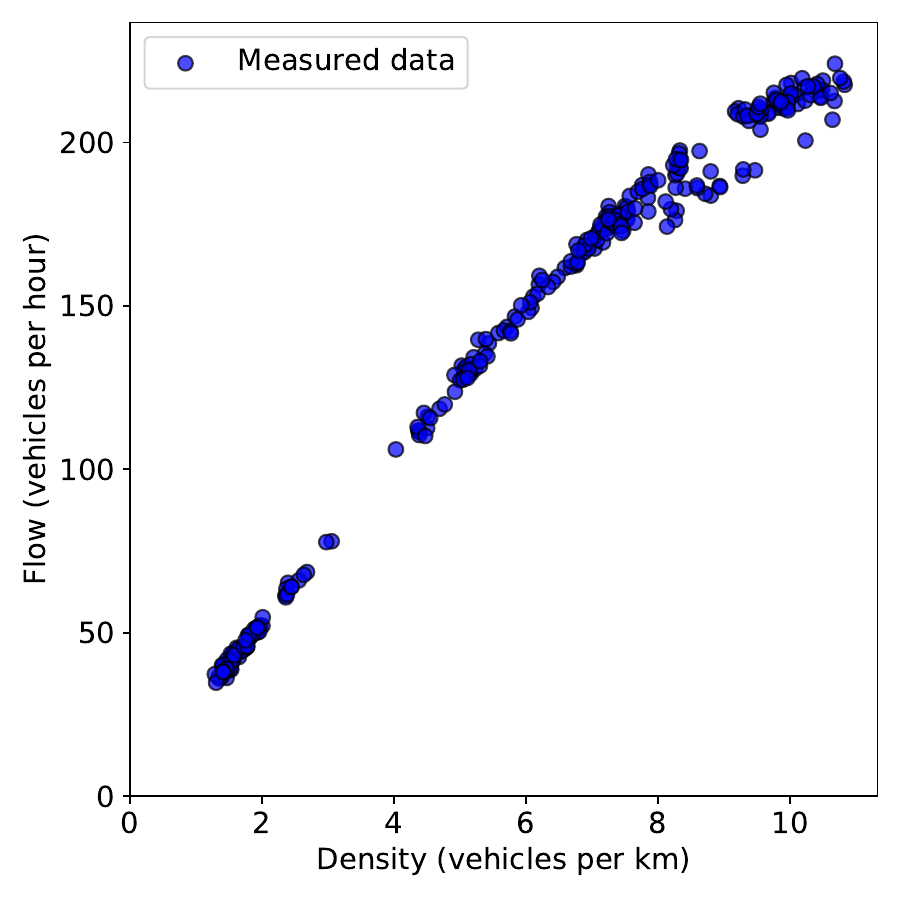}
    \caption{MFD for the simulation}
    \label{fig:sumo_mfd_24h}
\end{figure*}}

However, we also notice that the heterogeneity inside the city is quite high: we plot the density and flow for all measurements in the city from 3 am to 9 am in Figure~\ref{fig:mfd_3h_9h_network} as this period almost covers the period between the lowest and highest demand as presented in Figure~\ref{fig:sumo_lane_distribution_24h}. Note these are not MFDs but point measurements for each lane. However, we observe lanes are quite diverse and we can clearly see different capacities shown as red dashed horizontal lines in Figure~\ref{fig:mfd_3h_9h_network}.

\textbf{\begin{figure*}[!h]
    \centering
    \includegraphics[width=1\linewidth]{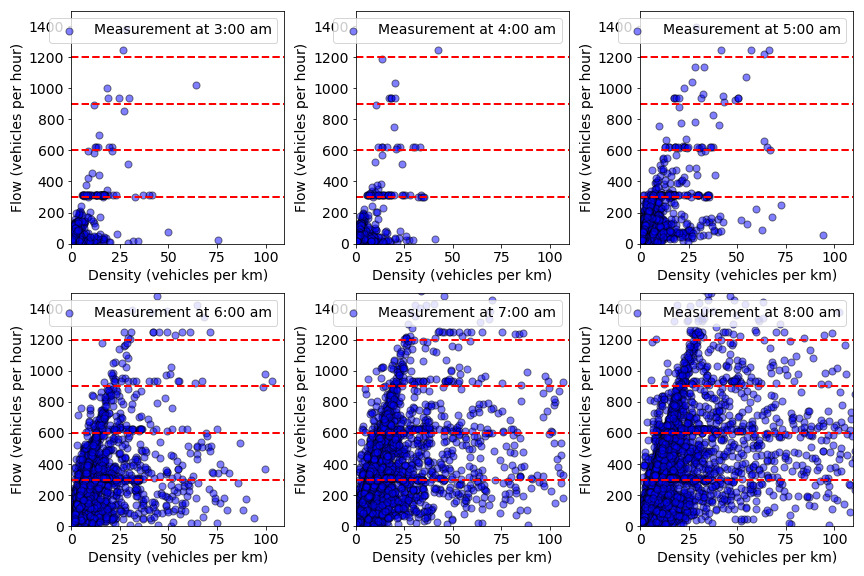}
    \caption{Measurements for the whole Ingolstadt from 3 am to 9 am}
    \label{fig:mfd_3h_9h_network}
\end{figure*}}

To tackle the heterogeneity existing in the whole network, we apply the K-means algorithm  using scikit-learn in Python~\cite{scikit-learn}.
Firstly, we calculate the average traffic states for each lane over the whole simulation process; 
secondly, we determine the optimal number of clusters using Elbow Method~\cite{thorndike1953}, which runs the K-means algorithm for different number of clusters, and plot corresponding within-cluster sum of square (WCSS). The optimal point is where the WCSS curve has the decreasing derivatives, which means the gain of increasing cluster number drops down. In our case, we select 4 as shown in the left of Figure~\ref{fig:clustering_4clusters}. This also fits our observations from Figure~\ref{fig:mfd_3h_9h_network} that there are roughly 4 capacities.
Finally, we plot the speed-density relations for these 4 clusters as shown in the right of Figure~\ref{fig:clustering_4clusters}. Apparently, cluster 1 is the low-class lanes with both low speeds and low densities and cluster 2 possesses the higher density and speed; cluster 3 might be located in the city center with high densities, while cluster 4 is with high speeds and low densities, which might be express way in the outskirt.

\textbf{\begin{figure*}[!h]
    \centering
    \includegraphics[width=1\linewidth]{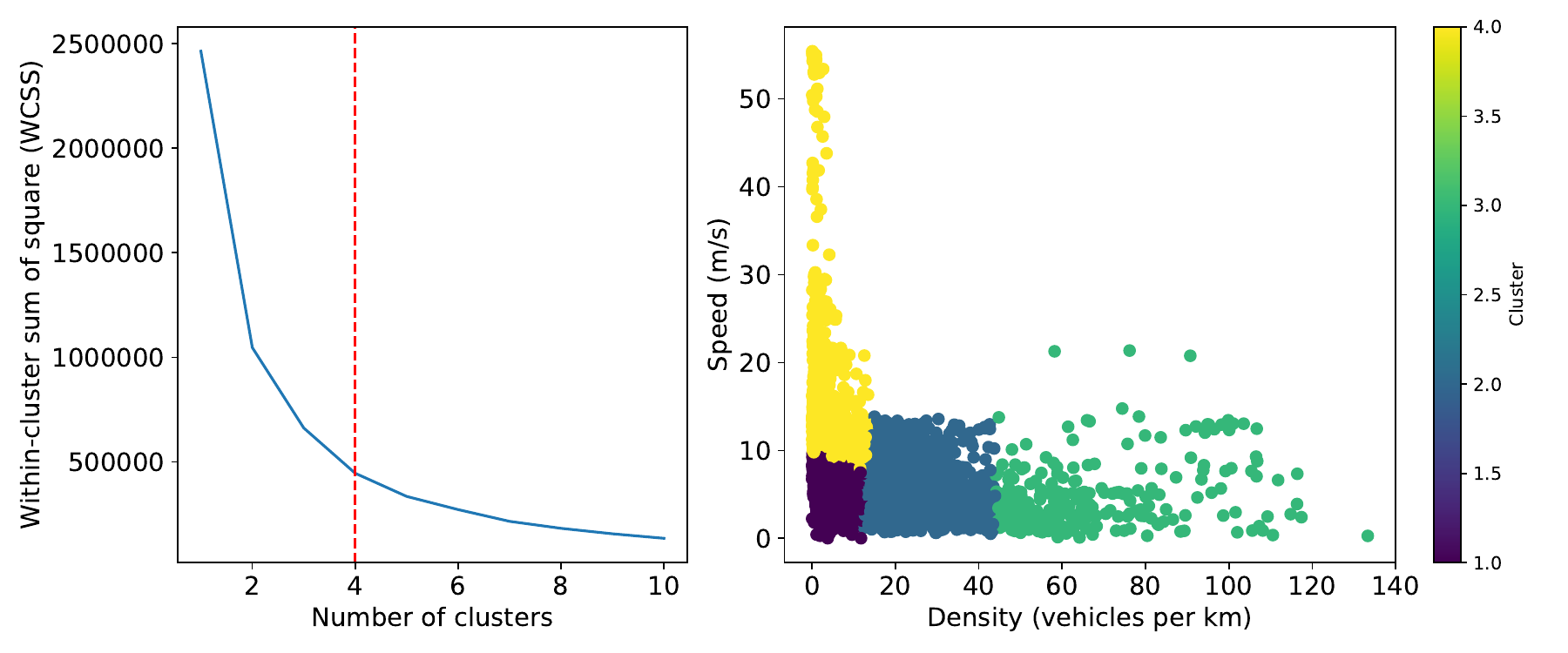}
    \caption{K-means clustering: Selection of the optimal number of clusters (left) and Speed-density measurements for different clusters (right)}
    \label{fig:clustering_4clusters}
\end{figure*}}

These conclusions can be proven through the MFDs for these 4 clusters as shown in Figure~\ref{fig:sumo_mfd_24h_4clusters}. Cluster 1 refers to all minor streets, while Cluster 2 and 3 shows a good fit of MFD, representing different capacities and road classes. Cluster 4 is rather low in density and flow though with a higher speed.

\textbf{\begin{figure*}[!h]
    \centering
    \includegraphics[width=1\linewidth]{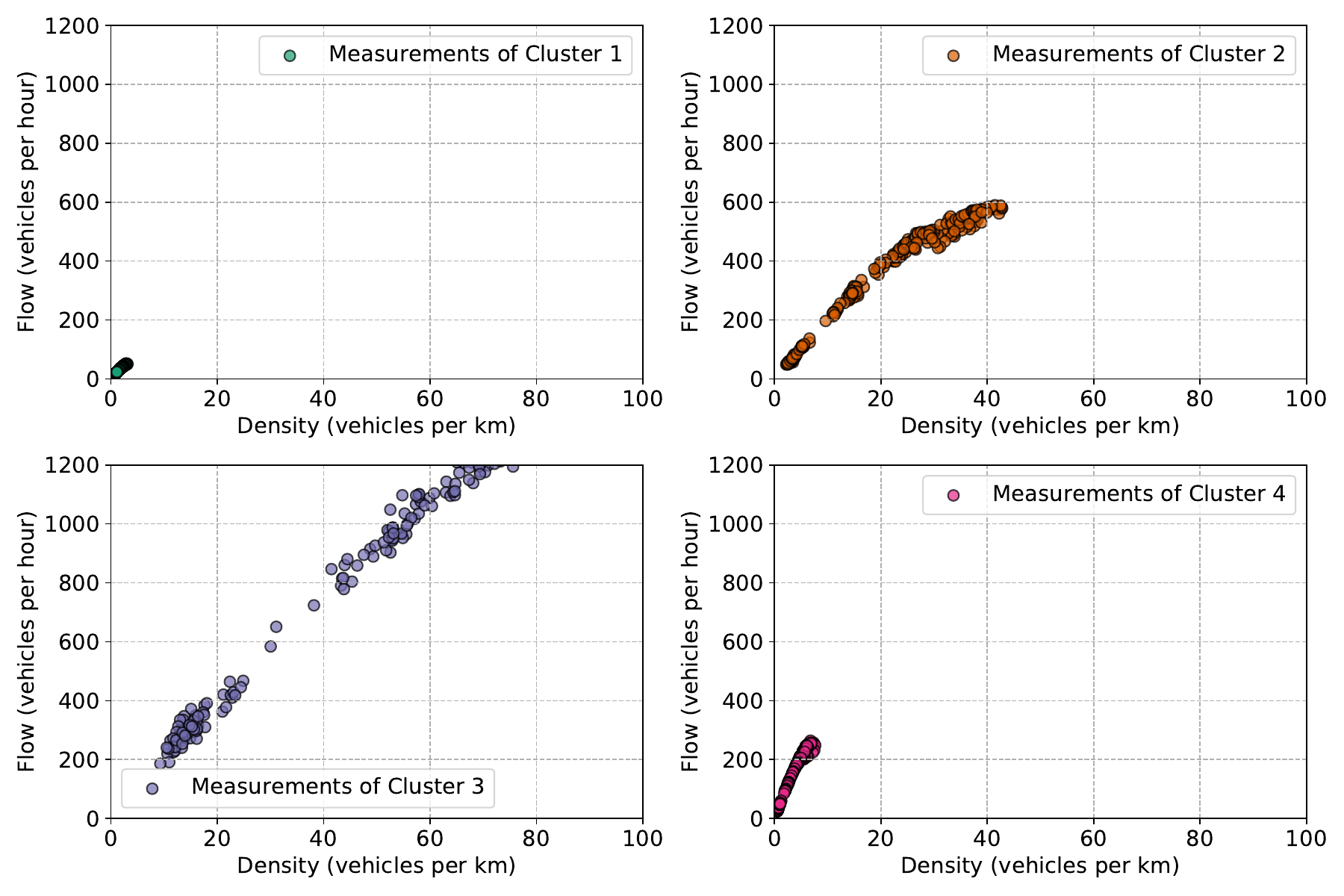}
    \caption{MFDs for the simulation for 4 clusters}
    \label{fig:sumo_mfd_24h_4clusters}
\end{figure*}}

\subsection{Results of Traffic State Estimation}

In the traffic state estimation, we do not use the 24-hour simulation because the dimension of data is too huge. We select the morning peak from 6 am to 11 am as the estimation period.
Firstly, we compare the estimated and measured densities and speeds based on 1,300 records. One record refers to one ego vehicle staying in the same lane. The comparison are visualized in Figure~\ref{fig:vaas_analysis_micro}. From the figure we can conclude that estimated densities do deviate a lot especially when the the densities are small, which again is due to the fact that it is hard for a moving or parking vehicle to capture a rather small density as already discussed in \cite{zhang2023novel}. However, the estimated vehicle speeds show a good fit.

\textbf{\begin{figure*}[!h]
    \centering
    \includegraphics[width=1\linewidth]{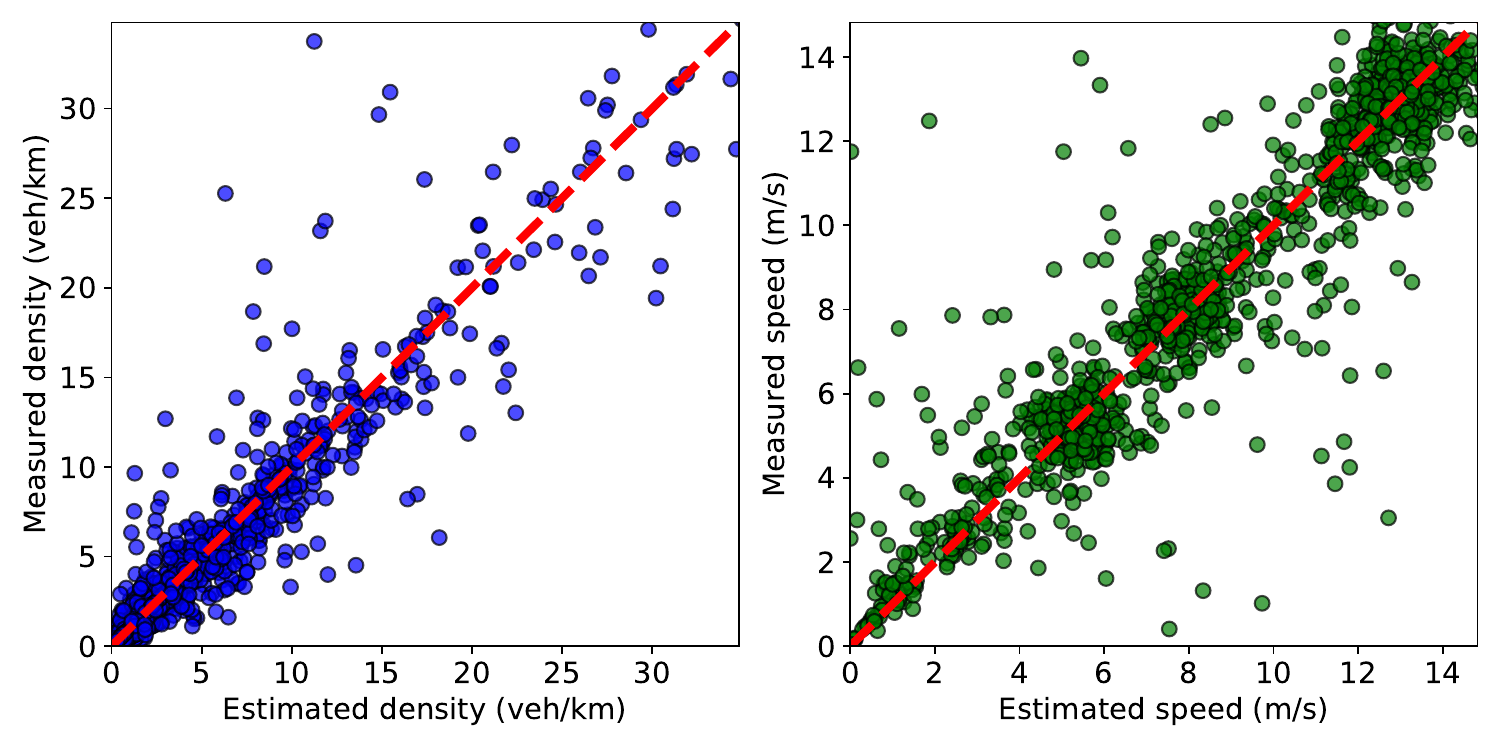}
    \caption{Microscopic (link-level) estimation of densities and speeds}
    \label{fig:vaas_analysis_micro}
\end{figure*}}

To further analyze how errors distribute at different scales, we plot the relative error (defined in Equation \ref{eq:relative_error}) of density and speed estimations in Figure \ref{fig:vaas_error}. As already talked, density estimation is not precise when the overall density is low, while the speeds' estimation always have the similar relative error regardless of the measured speed. However, the overall relative errors of both estimations are not huge. 
Generally, densities are overestimated because the moving observers will anyway count itself into the density calculation, which can be biased. However, this will not be a problem for parking observers.

\begin{equation}\label{eq:relative_error}
    \text{Relative Error} = \frac{| \text{True Value} - \text{Estimated Value} |}{| \text{True Value} |}
\end{equation}

\textbf{\begin{figure*}[!h]
    \centering
    \includegraphics[width=1\linewidth]{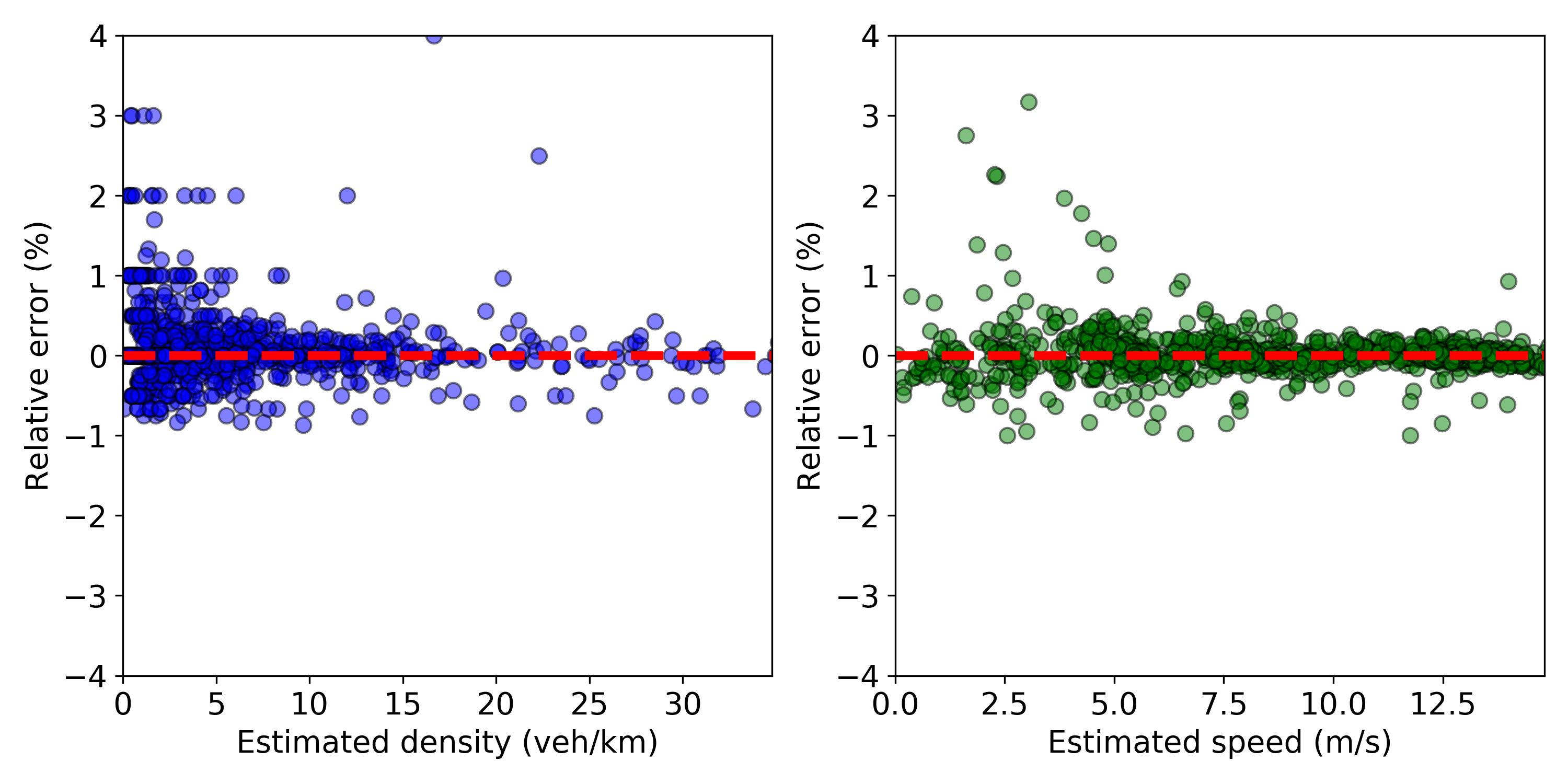}
    \caption{Error distribution of densities and speeds}
    \label{fig:vaas_error}
\end{figure*}}

Starting from link level, we aggregate the traffic states firstly at the cluster level in Figure~\ref{fig:mfd_est_5h_4clusters}. All 4 clusters shows the underestimation of density, and thus flow. This could due to the fact that even within the clusters, different lanes are diverse in density and speed profile.
The same phenomenon appears if we further aggregate the traffic state estimation to the network level in Figure~\ref{fig:sumo_mfd_5h_comparison}. Note that we present speed and density relation here because the estimated flow is way smaller than the measurements. From the Figure we can clearly observe that densities are even more underestimated. This leads to further research on how to increase the homogeneity within the cluster or even increase the penetration rate of AVaaS for further improvements.

\textbf{\begin{figure*}[!h]
    \centering
    \includegraphics[width=1\linewidth]{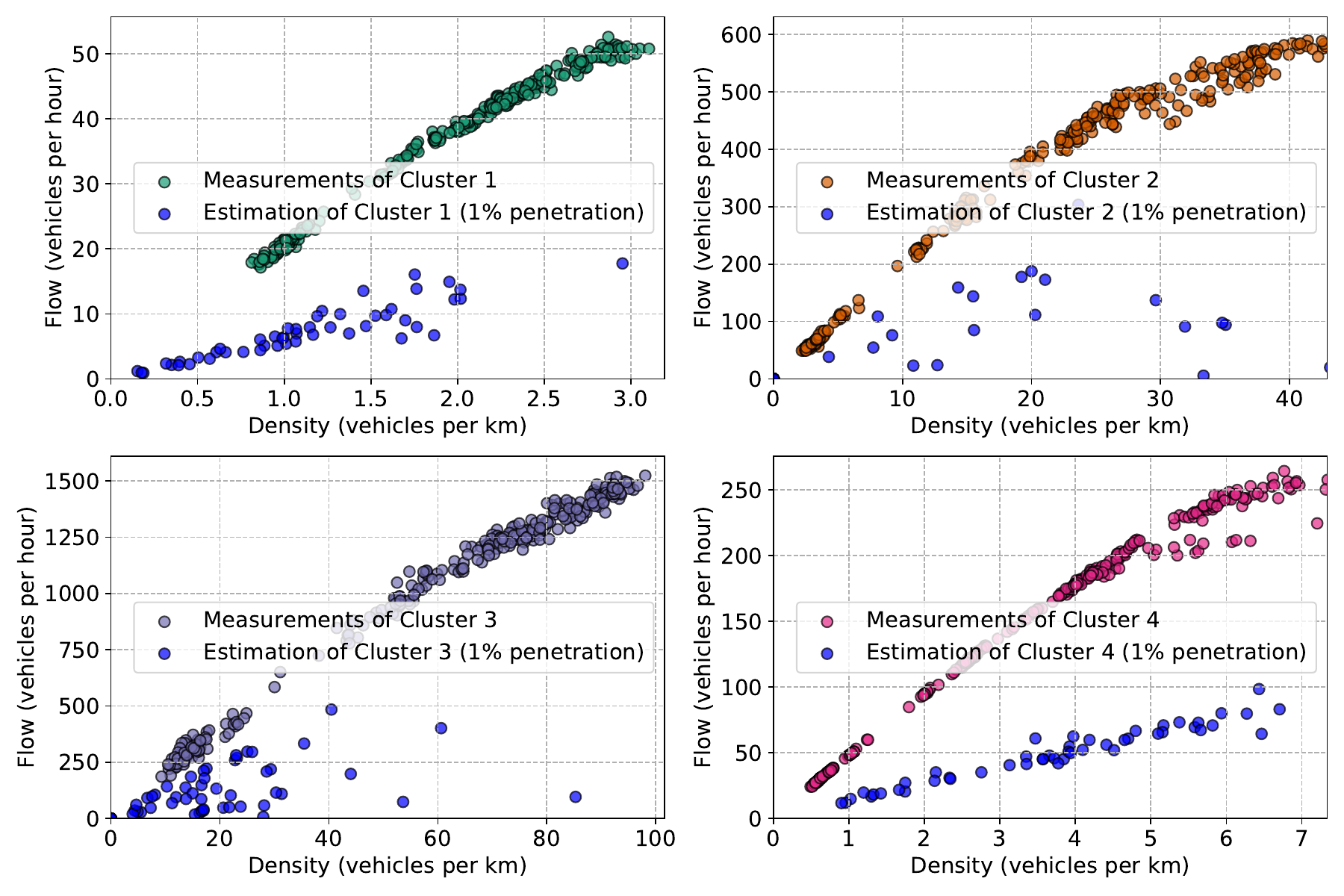}
    \caption{Estimation of MFDs for different clusters}
    \label{fig:mfd_est_5h_4clusters}
\end{figure*}}

\textbf{\begin{figure*}[!h]
    \centering
    \includegraphics[width=0.6\linewidth]{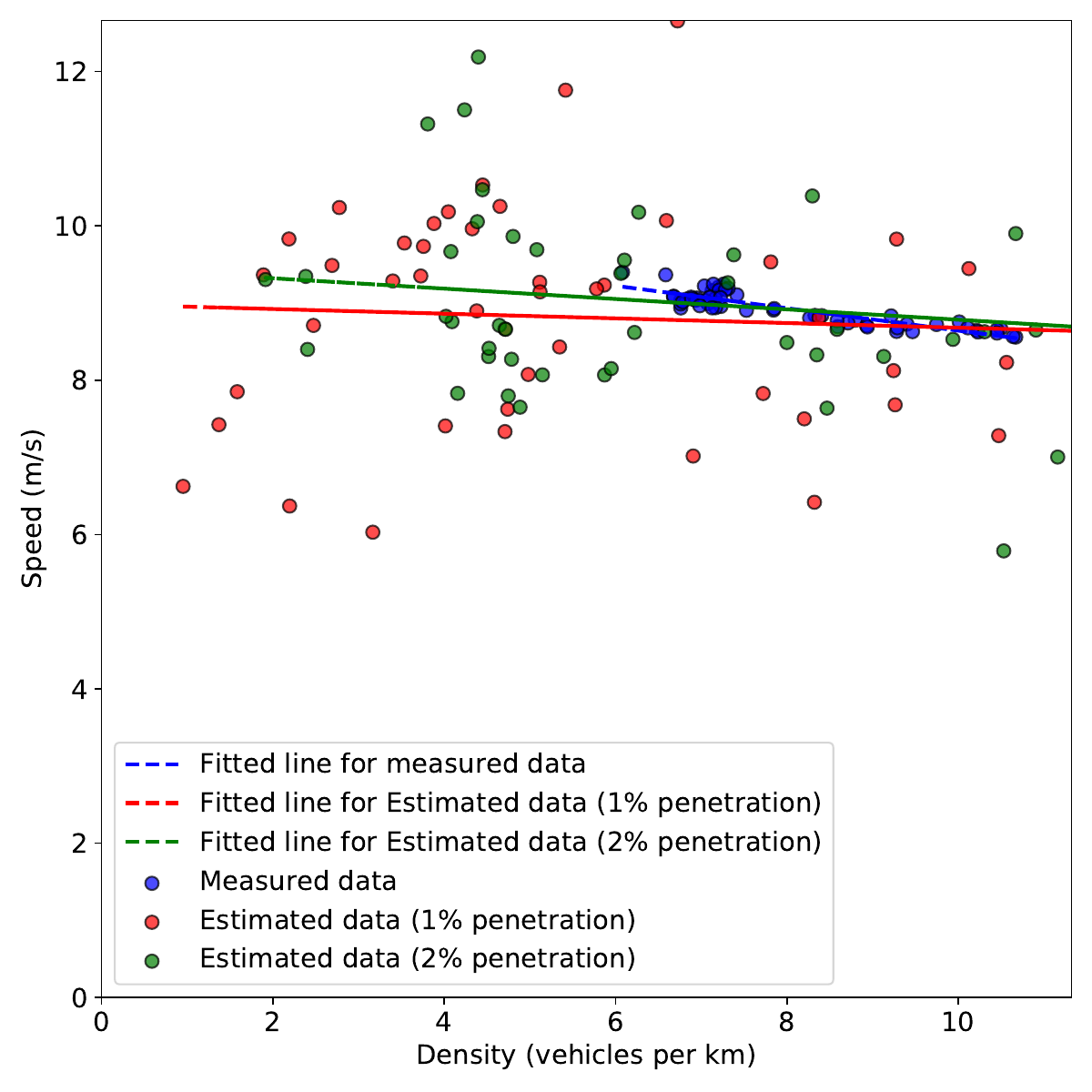}
    \caption{Comparison of MFDs based on measured and estimated data}
    \label{fig:sumo_mfd_5h_comparison}
\end{figure*}}

\section{Conclusion}

In this paper, we expand the dimension of Autonomous Vehicles as a Sensor (AVaaS) from a concept to a real-world network study. Our contributions can be summarized as three folds: firstly, we improve the data simulation process by incorporating real-world detection factors including range and attributes of sensors, resulting in more accurate and reliable traffic data collection; secondly, we demonstrate the effectiveness of our proposed methods through a real-world network case study, showcasing the practical applicability of our approach in real-life scenarios; lastly, we present the traffic state estimation for link level, cluster level, and network level.

However, we also have some limitations.
For the perception, we still model it through distance and detection zone-based methods, which do not include occlusion in reality.
The second limitation is that our method does not perform well in the cluster and network level, indicating a more dynamic clustering method based on real-time AVaaS concept is required.
Lastly, we do not include the error model because the complexity will increase dramatically as stated in \cite{pechinger2023roadside}.

To tackle these limitations, we have planned some future works.
Regarding the perception, we plan to implement a perception method using Floating Car Observers (FCOs) in the microscopic simulations~\cite{gerner2023enhancing}. In this paper, authors convert two-dimensional simulation scenarios into three-dimension environment and use neural network for computer vision (CV) to detect the surrounding traffic participants. Based on that, we do not work on the CV algorithm but use them as a more realistic way to detect traffic environment and estimate the traffic states accordingly.
For cluster- and network-level perception, we plan to develop a more reliable clustering and network representation methods based on AVaaS with a relatively low penetration rate, which is the focus of further applications in this field.

\section*{Acknowledgments}
This work was funded by the German Federal Ministry for Economic Affairs and Climate Action (BMWK) and by the European Union in the frame of NextGenerationEU within the project STADT:up (FKZ 19A22006T). [The authors would like to thank the consortium for the successful cooperation.]

\bibliographystyle{unsrt}  
\bibliography{references}

\end{document}